\documentclass[prb,twocolumn,showpacs]{revtex4}
\usepackage{amsmath}
\usepackage{amssymb}
\usepackage{graphicx}
\usepackage{bm}
\usepackage{dcolumn}

\newcommand{\bpm}{\begin{pmatrix}}
\newcommand{\epm}{\end{pmatrix}}

\begin{document}

\title{Properties of superconductivity on the density wave background with
small ungapped Fermi surface pockets.}
\author{P. D. Grigoriev}
\email{grigorev@itp.ac.ru}
\affiliation{L. D. Landau Institute for Theoretical Physics, Chernogolovka, Russia}
\altaffiliation[Visiting address: \ ]{Institut Laue Langevin, Grenoble, France}
\date{\today }

\begin{abstract}
We investigate the properties and the microscopic structure of
superconductivity (SC), coexisting and sharing the common conducting band
with density wave (DW). Such coexistence may take place when the nesting of
the Fermi surface (FS) is not perfect, and in the DW state some
quasi-particle states remain on the Fermi level and lead to the Cooper
instability. The dispersion of such quasi-particle states is, in general,
very different from that without DW. Therefore, the properties of SC on the
DW background may strongly differ from those without DW. The upper critical
field $H_{c2}$ in such a SC state increases as the system approaches the
critical pressure, where the ungapped quasi-particles and superconductivity
just appear, and it may considerably exceed the usual $H_{c2}$ value without
DW. The SDW background strongly suppresses the singlet SC pairing, while it
does not affect so much the triplet SC transition temperature. The results
obtained explain the experimental observations in layered organic metals
(TMTSF)$_{2}$PF$_{6}$ and $\alpha $-(BEDT-TTF)$_{2}$KHg(SCN)$_{4}$, where SC
appears in the DW states under pressure and shows many unusual properties.
\end{abstract}

\maketitle

\section{Introduction}

The interplay between superconductivity (SC) and insulating charge or spin
density wave states is a subject of an active investigation for more than 30
years (for a review see, e.g., Ref. [\onlinecite{Review1}]). The density
wave (DW) is traditionally considered as a strong obstacle for the formation
of SC, because it creates an energy gap on the Fermi level.\cite%
{Levin,Balseiro,Milans} The coexistence of DW and SC has been considered in
metals with several conducting bands or with imperfect nesting, when even in
the DW state there is a finite electron density on the Fermi level.\cite%
{Bilbro,Machida,Psaltakis} Then the transition temperature $T_{c}^{SC}$ to
the SC state reduces exponentially when the DW is formed, because the
electrons, participating in the formation of DW, drop out from the SC
condensate.\cite{Bilbro,Machida}

However, in several compounds [e.g., in layered organic superconductors
(TMTSF)$_{2}$PF$_{6}$ and $\alpha $-(BEDT-TTF)$_{2}$KHg(SCN)$_{4}$],\cite%
{Vuletic,CDWSC} the SC transition temperature on the DW background is very
close to (or even exceeds) $T_{c}^{SC}$ without DW. In (TMTSF)$_{2}$PF$_{6}$
superconductivity coexists with spin-density-wave (SDW) state at temperature
below $T_{c}^{SC}\approx 1.1K$ in the pressure interval above some critical
pressure $P_{c1}\approx 8.5$kbar, but below $P_{c}\approx 9.5$kbar, at which
the SDW phase undergoes the 1st order phase transition into metallic state.%
\cite{Vuletic} This fact is even more surprising, because this compound has
only one quasi-one-dimensional (Q1D) conducting band. A special attention
was given to the fact, that the upper critical field $H_{c2}$ in this
superconducting state exceeds several times the expected paramagnetic limit,%
\cite{Clogston} (see Refs. [\onlinecite{LeeTripletMany,LeeAngularHc}]), and
no change in the Knight shift has been observed in this compound as
temperature lowers to this SC state\cite{LeeKnightShift}. Both these
features suggest the spin-triplet superconducting paring in (TMTSF)$_{2}$PF$%
_{6}$. In addition, the upper critical field $H_{c2}$ perpendicular to the
conducting layers strongly increases as pressure approaches $P_{c1}$ and has
an unusual upward curvature as function of temperature\cite{Hc2Pressure},
suggesting that the SDW has a very strong influence on the SC properties of
this phase. The electronic structure of this mixed phase is still under
debates. A phase separation in a form of macroscopic metal and DW domains,%
\cite{Vuletic,Hc2Pressure} being natural with the constant volume
constraint, seems strange at fixed pressure, when the whole sample may
choose the state with the lowest free energy. The pressure and temperature
dependence of the upper critical field requires,\cite{Hc2Pressure} that the
size $d$ of the SC domains, if they exist, must be much less than the SC
coherence length $\xi _{SC}\sim 10^{-4}cm$ as pressure approaches $P_{c1}$
[see Eq. (\ref{Hslab}) and the discussion in Sec. III]. This raises many
questions about the structure of such a mixed state, because if the domain
width is comparable to the SDW coherence length, this confinement of the
electron wave functions costs additional energy greater than the SC energy
gap. The angular magnetoresistance oscillations\cite{LeeBrown} do not give a
definite test whether the spatial phase separation occurs on a scale greater
than the SC coherence length.

An alternative to the picture \cite{Vuletic,Hc2Pressure} of macroscopic DW
and normal (or SC) domains in (TMTSF)$_{2}$PF$_{6}$ has been proposed
recently.\cite{GG,GGPRB2007} According to Ref. \cite{GGPRB2007}, there are
two different structures where SC coexists microscopically with DW. In the
first structure, the destruction of the insulating DW phase at $P>P_{c1}$
goes via forming the ungapped metallic pockets in the electronic spectrum,
that spread over the momentum space, merging into the normal metallic state
gradually or via a phase transition. This scenario, looking similar to the
one studied in Refs. [\onlinecite{Bilbro,Machida}], however, differs from
it, because the formation of DW strongly modifies the quasi-particle
dispersion in the ungapped parts of the Fermi surface, changing the
properties of SC state on the DW background. In the second scenario, the DW
order parameter at pressure $P>P_{c1}$ becomes spatially nonuniform by means
of amplitude solitons. These soliton structures are familiar in
charge-density-wave (CDW) states at high pressure or in magnetic field (see,
e.g., reviews in Refs. [\onlinecite{BrazKirovaReview}],[\onlinecite{SuReview}%
]). The normal or SC phase appears first as the metallic domain walls, and
the concentration of these soliton walls increases with increasing pressure.
At finite density of solitons, i.e. above $P_{c1},$ the electron wave
functions of single solitons strongly overlap, forming a new periodic
conducting metallic network on the DW background. If it were not for the 1st
order transition with the further increase of soliton density, the new phase
is expected to merge gradually into the normal state.

Both microscopic structures may appear in DW superconductors. Nuclear
magnetic resonance (NMR) experiments\cite{Brown} are consistent with the
scenario, where the phase separation takes place on the microscopic scale
not exceeding the DW coherence length, thus supporting either of the above
scenarios. In both scenarios, at low enough temperature, superconductivity
appears at pressure $P>P_{c1}$,\cite{GGPRB2007} and the DW have a strong
influence on the properties of such a mixed SC state. In particular, the SDW
background suppresses the spin-singlet SC pairing, making spin-triplet
pairing to be expected\cite{GGPRB2007} in agreement with experiments in
(TMTSF)$_{2}$PF$_{6}$.\cite{LeeTripletMany,LeeAngularHc,LeeKnightShift} This
feature appears due to the spin-dependent scattering on the SDW condensate,
and it does not happen when superconductivity coexists with CDW.

In the present paper we follow the ideas of Ref. [\onlinecite{GGPRB2007}]
and study in detail the microscopic structure and properties of the mixed
SC-DW state in the first scenario of uniform DW with ungapped metallic
pockets above $P>P_{c1}$. In Sec. II we generalize the model of Ref. [%
\onlinecite{GGPRB2007}] to the case of more realistic e-e interaction with
backward and forward scattering, and describe in detail the uniform DW state
with ungapped pockets. In Sec. III we estimate the SC transition temperature
and the upper critical field $H_{c2}$ in the DW-SC mixed state, and show,
that $H_{c2}$ strongly increases as pressure approaches the critical value $%
P_{c1}$, in agreement with experiments in (TMTSF)$_{2}$PF$_{6}$ and $\alpha $%
-(BEDT-TTF)$_{2}$KHg(SCN)$_{4}$. In Sec. IV we study SC on SDW background
and show that SDW suppresses the spin-singlet SC ordering. Our results are
aimed mainly to quasi-1D metals, but also can be applied to other DW
superconductors with slightly imperfect nesting.

\section{The model and the DW state without superconductivity}

In quasi-1D metals the free electron dispersion without magnetic field
writes down as
\begin{equation}
\varepsilon (\boldsymbol{k})=\pm v_{F}(k_{x}\mp k_{F})+t_{\perp }(\mathbf{k}%
_{\perp }).  \label{1}
\end{equation}%
The electron dispersion in the easy-conducting (chain) $x$-direction is
strong and can be linearized near the Fermi surface (FS). The interchain
dispersion $t_{\perp }({\boldsymbol{k}}_{\perp })$ is much weaker and given
by the tight-binding model with few leading terms:
\begin{equation}
t_{\perp }({\boldsymbol{k}}_{\perp })=2t_{b}\cos (k_{y}b)+2t_{b}^{\prime
}\cos (2k_{y}b)+2t_{c}\cos (k_{z}\tilde{c}),  \label{dispersion}
\end{equation}%
where $b,\tilde{c}$ are the lattice constants in the $y$- and $z$-direction.
The dispersion along the $z$-axis is considerably weaker than along the $y$%
-direction and does not play any role in the analysis below. The FS consists
of two warped sheets and possesses an approximate nesting property, $%
\varepsilon (\boldsymbol{k})\approx -\varepsilon (\boldsymbol{k}\mathbf{-}%
\boldsymbol{Q})$, which leads to the formation of DW at low temperature. The
nesting property is only spoiled by the second term $t_{b}^{\prime }(k_{y}%
\mathbf{)}$ in Eq. (\ref{dispersion}), which, therefore, is called the
"antinesting" term. Increase of the latter with applied pressure leads to
the transition in the gapped DW state at $P=P_{c1}$, where the ungapped
pockets on the FS or isolated soliton walls\cite{BGL} first appear. In the
pressure interval $P_{c1}<P<P_{c}$ the new state develops, where the DW
coexists with superconductivity at rather low temperature $T<T_{c}^{SC}$,
while at high temperature $T>T_{SC}$ the DW state coexists with the normal
metal phase.

The electron Hamiltonian is
\begin{equation}
\hat{H}=\hat{H}_{0}+\hat{H}_{\mathrm{int}},  \label{H}
\end{equation}%
with the free-electron part in momentum representation
\begin{equation}
\hat{H}_{0}=\sum_{{\boldsymbol{k}}}\varepsilon (\boldsymbol{k})a_{\alpha
}^{\dag }({\boldsymbol{k}})a_{\alpha }({\boldsymbol{k}})  \label{H0}
\end{equation}%
and the interaction part
\begin{eqnarray}
\hat{H}_{\mathrm{int}} &=&\frac{1}{2}\sum_{{\boldsymbol{k}}{\boldsymbol{k}}%
^{\prime }{\boldsymbol{Q}}}V_{\alpha \beta \gamma \delta }({\boldsymbol{Q}}%
)a_{\alpha }^{\dag }({\boldsymbol{k}}+{\boldsymbol{Q}})a_{\beta }({%
\boldsymbol{k}})  \notag \\
&&\times a_{\gamma }^{\dag }({\boldsymbol{k}}^{\prime }-{\boldsymbol{Q}}%
)a_{\delta }({\boldsymbol{k}}^{\prime }).  \label{Hint0}
\end{eqnarray}%
Here and below we imply the summation over repeating spin indices. The
interaction potential is
\begin{equation}
V_{\alpha \beta \gamma \delta }({\boldsymbol{Q}})=U_{c}\left( \boldsymbol{Q}%
\right) -U_{s}\left( \boldsymbol{Q}\right) \vec{\sigma}_{\alpha \beta }\vec{%
\sigma}_{\gamma \delta },  \label{vQ}
\end{equation}%
where $\vec{\sigma}_{\alpha \beta }$ are the Pauli matrices. For the
formation of DW, the value of this potential at the nesting vector ${%
\boldsymbol{Q=Q}}_{N}$ is only important. The values $U_{c}\left(
\boldsymbol{Q}_{N}\right) $ and $U_{s}\left( \boldsymbol{Q}_{N}\right) $ are
called the charge and spin coupling constants. Depending on their ratio, the
charge or spin density wave is formed.

For superconducting pairing, both the momentum and frequency dependence of
the potential (\ref{vQ}) is important, being different for different
compounds. Below we consider only a simplified model, similar to the BCS
model,\cite{BCS} where the frequency dependence of the interaction potential
is taken into account only through the ultraviolet cutoff (Debye frequency)
in the Cooper loop. The phonon-mediated electron pairing produces only the
spin-independent charge coupling $U_{c}\left( \boldsymbol{Q}\right) $, and
in the study of superconductivity we put $U_{s}\left( \boldsymbol{Q}\right)
=0$. As concerns the momentum dependence of $U_{c}\left( \boldsymbol{Q}%
\right) $, in 1D and quasi-1D metals one, usually, distinguishes only the
backward and forward scattering:
\begin{equation}
U_{c}({\boldsymbol{Q}})=\left\{
\begin{array}{c}
U_{c}^{f},~Q_{x}\ll 2k_{F} \\
U_{c}^{b},~Q_{x}\approx 2k_{F}%
\end{array}%
\right. .  \label{vQ1}
\end{equation}%
Depending on the signs and the ratio of backward $U_{c}^{b}$ and forward $%
U_{c}^{f}$ coupling constants, one has singlet or triplet SC pairing. The
Hamiltonian (\ref{H}) does not include the spin-orbit interaction, which is
assumed to be weak.

Below we assume the DW transition temperature to be much greater than the SC
transition temperature, $T_{c}^{DW}\gg T_{c}^{SC}$, which corresponds to
most DW superconductors. For example, in (TMTSF)$_{2}$PF$_{6}$ $%
T_{c}^{SDW}\approx 8.5K\gg T_{c}^{SC}\approx 1.1K$, and in $\alpha $%
-(BEDT-TTF)$_{2}$KHg(SCN)$_{4}$, $T_{c}^{CDW}\approx 8K\gg T_{c}^{SC}\approx
0.1K$. Therefore, we first study the structure of the DW state in the
pressure interval $P_{c1}<P<P_{c}$, and then consider the superconductivity
on this background.

\subsection{The uniform DW state with ungapped states}

In the case of the uniform DW order parameter, $\Delta _{0}(x)=\Delta
_{0}=const(T,P)$, the electron Green functions in the DW state in the
mean-field approximation can be written down explicitly. We introduce the
thermodynamic Green function
\begin{equation}
\hat{g}_{\alpha \beta }({\boldsymbol{k}}^{\prime },{\boldsymbol{k}},\tau
-\tau ^{\prime })=\langle T_{\tau }\{a_{\alpha }^{\dag }({\boldsymbol{k}}%
^{\prime },\tau ^{\prime })a_{\beta }({\boldsymbol{k}},\tau )\}\rangle ,
\label{Gkk}
\end{equation}%
where the operators are taken in the Heisenberg representation, and the
Green function $\hat{g}_{\alpha \beta }({\boldsymbol{k}}^{\prime },{%
\boldsymbol{k}},\tau -\tau ^{\prime })$ is an operator in the spin space.
The CDW order parameter%
\begin{equation}
\hat{\Delta}_{{\boldsymbol{Q}}}=U_{c}\sum_{{\boldsymbol{k}}}\hat{g}({%
\boldsymbol{k}}-{\boldsymbol{Q}},{\boldsymbol{k}},-0)=\Delta _{{\boldsymbol{Q%
}}}  \label{DeltaCDW}
\end{equation}%
is a unity operator in spin space, and the SDW order parameter is
\begin{equation}
\hat{\Delta}_{{\boldsymbol{Q\alpha \beta }}}=U_{s}\left( \vec{\sigma}%
_{\alpha \beta }\cdot \vec{\sigma}_{\gamma \delta }\right) \sum_{{%
\boldsymbol{k}}}\hat{g}_{\gamma \delta }({\boldsymbol{k}}-{\boldsymbol{Q}},{%
\boldsymbol{k}},-0)=(\vec{\hat{\sigma}}\vec{l})\Delta _{\boldsymbol{Q}},
\label{DeltaQs}
\end{equation}%
where the complex vector $\vec{l}$ determines the polarization of the SDW.
In the presence of magnetic field $\vec{H}$ and without internal magnetic
anisotropy, $\vec{l}\perp \vec{H}$. Below the external magnetic field is
taken to be rather weak to only affect SC but not the DW,\cite%
{CommentStrongH} because strong magnetic field would suppress SC. We
consider only one DW order parameter, i.e. $\Delta _{\boldsymbol{Q}}\neq 0$
only for $\boldsymbol{Q}=\pm {\boldsymbol{Q}}_{N}$, where ${\boldsymbol{Q}}%
_{N}\approx 2k_{F}{\boldsymbol{e}}_{x}+(\pi /b){\boldsymbol{e}}_{y}+(\pi /%
\tilde{c}){\boldsymbol{e}}_{z}$, and ${\boldsymbol{e}}_{x},{\boldsymbol{e}}%
_{y},{\boldsymbol{e}}_{z}$ are the unit vectors in $x,y,z$ directions. In
the mean-field approximation one has
\begin{equation*}
\hat{H}_{\mathrm{int}}=\frac{1}{2}\sum_{{\boldsymbol{Q}}{\boldsymbol{k}}%
}a_{\alpha }^{\dag }({\boldsymbol{k}}+{\boldsymbol{Q}})a_{\beta }({%
\boldsymbol{k}})\hat{\Delta}_{{\boldsymbol{Q}}\alpha \beta }.
\end{equation*}%
Hermicity of the Hamiltonian requires $\hat{\Delta}_{-{\boldsymbol{Q}}\alpha
\beta }=\hat{\Delta}_{{\boldsymbol{Q}}\beta \alpha }^{\ast }$. Below we omit
the explicit spin indices, keeping only the "hat" symbol above the spin
operators. For SDW the equations of motion in the frequency representation
are
\begin{equation}
\lbrack i\omega -\varepsilon (\mathbf{k})]\hat{g}({\boldsymbol{k}}^{\prime },%
{\boldsymbol{k}},\omega )-\sum_{\boldsymbol{Q}}\Delta _{0}(\vec{\hat{\sigma}}%
\vec{l})\hat{g}({\boldsymbol{k}}^{\prime },{\boldsymbol{k}}-{\boldsymbol{Q}}%
,\omega )=\delta _{{\boldsymbol{k}}^{\prime }{\boldsymbol{k}}}  \label{GEw}
\end{equation}%
If we neglect the scattering into the states with $|k_{x}|\gtrsim 2k_{F}$,
the equations (\ref{GEw}) decouple:%
\begin{equation}
\left(
\begin{array}{cc}
i\omega _{n}-\varepsilon (\mathbf{k}) & \Delta _{0}(\vec{\hat{\sigma}}\vec{l}%
) \\
\Delta _{0}^{\ast }(\vec{\hat{\sigma}}\vec{l}) & i\omega _{n}-\varepsilon ({%
\boldsymbol{k}}-{\boldsymbol{Q}})%
\end{array}%
\right) \hat{G}=\hat{I},  \label{ginv}
\end{equation}%
where the matrix Green function%
\begin{equation}
\hat{G}\equiv \left(
\begin{array}{cc}
g^{RR}({\boldsymbol{k}},{\boldsymbol{k}},\omega ) & g^{LR}({\boldsymbol{k}}-{%
\boldsymbol{Q}},{\boldsymbol{k}},\omega )(\vec{\hat{\sigma}}\vec{l}) \\
g^{RL}({\boldsymbol{k}},{\boldsymbol{k}}-{\boldsymbol{Q}},\omega )(\vec{\hat{%
\sigma}}\vec{l}) & g^{LL}({\boldsymbol{k}}-{\boldsymbol{Q}},{\boldsymbol{k}}-%
{\boldsymbol{Q}},\omega )%
\end{array}%
\right) ,  \label{hatG}
\end{equation}%
$\hat{I}$ is the $2\times 2$ identity matrix, and the $R$ and $L$
superscripts denote the right and left FS sheet of electrons:%
\begin{equation}
a_{\alpha }({\boldsymbol{k}},\tau )\equiv \left\{
\begin{array}{c}
a_{\alpha }^{R}({\boldsymbol{k}},\tau ),~k_{x}>0 \\
a_{\alpha }^{L}({\boldsymbol{k}},\tau ),~k_{x}<0%
\end{array}%
\right. .  \label{aLR}
\end{equation}%
The electron Green functions in the CDW state are obtained from Eqs. (\ref%
{ginv}),(\ref{hatG}) by removing the spin factor $(\vec{\hat{\sigma}}\vec{l}%
) $ from the nondiagonal elements.

Introducing the notations
\begin{equation}
\varepsilon _{\pm }({\boldsymbol{k}}^{\prime },{\boldsymbol{k}})=\left[
\varepsilon ({\boldsymbol{k}}^{\prime })\pm \varepsilon (\mathbf{k})\right]
/2  \label{epspm}
\end{equation}%
and%
\begin{equation}
E_{1,2}\left( \mathbf{k}\right) \equiv \varepsilon _{+}({\boldsymbol{k}},{%
\boldsymbol{k}}-{\boldsymbol{Q}})\pm \sqrt{\varepsilon _{-}^{2}({\boldsymbol{%
k}},{\boldsymbol{k}}-{\boldsymbol{Q}})+|\Delta _{0}|^{2}},  \label{E1}
\end{equation}%
from (\ref{ginv}) one has%
\begin{equation}
g^{LR}({\boldsymbol{k}}-{\boldsymbol{Q}},{\boldsymbol{k}},\omega )=\frac{%
\Delta _{0}}{[i\omega -E_{1}\left( \mathbf{k}\right) ][i\omega -E_{2}\left(
\mathbf{k}\right) ]},  \label{GLR}
\end{equation}%
\begin{equation*}
g^{RL}({\boldsymbol{k}},{\boldsymbol{k}}-{\boldsymbol{Q}},\omega )=\frac{%
\Delta _{0}^{\ast }}{[i\omega -E_{1}\left( \mathbf{k}\right) ][i\omega
-E_{2}\left( \mathbf{k}\right) ]},
\end{equation*}%
and%
\begin{equation}
g^{RR}({\boldsymbol{k}},{\boldsymbol{k}},\omega )=\frac{i\omega -\varepsilon
(\mathbf{k})}{[i\omega -E_{1}\left( \mathbf{k}\right) ][i\omega -E_{2}\left(
\mathbf{k}\right) ]}=g^{LL}({\boldsymbol{k}},{\boldsymbol{k}},\omega ).
\label{GRR}
\end{equation}

\begin{figure}[tb]
\includegraphics[width=0.49\textwidth]{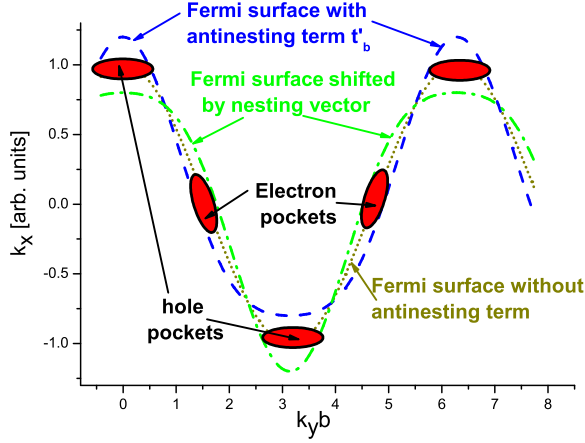}
\caption{(Color online) The schematic picture of small open pockets on one
Fermi surface sheet, which get formed when the antinesting term $\protect%
\varepsilon _{+}$ in Eq. (\protect\ref{E1})\ exceeds the DW energy gap $%
\Delta _{0}$. The blue dashed line shows the Fermi surface sheet with
imperfect nesting, i.e. with $2t_{b}^{\prime }>\Delta _{0}$. The green
dash-dotted line shows the other Fermi surface sheet shifted by the nesting
vector. If the nesting was perfect, these two lines would coincide. The
dotted brown line shows the perfectly nested Fermi surface sheet. The red
solid ellipses are the small Fermi surface pockets, that appear in the DW
state when $2t_{b}^{\prime }>\Delta _{0}$, i.e. when pressure exceeds $%
P_{c1} $.}
\label{FigPockets}
\end{figure}

The ungapped pockets on the Fermi surface with energy spectrum (\ref{E1})
appear when $\left\vert \varepsilon _{+}({\boldsymbol{k}})\right\vert _{\max
}=2t_{b}^{\prime }>|\Delta _{0}|$, and these pockets are responsible for the
Cooper instability at $P>P_{c1}$. With the tight-binding dispersion (\ref%
{dispersion}) at $P>P_{c1}$ there are four ungapped pockets on each of the
two sheets of the original Fermi surface: two electron pockets with $%
E_{2}\left( k\right) =\varepsilon _{+}({\boldsymbol{k}})+\sqrt{\varepsilon
_{-}^{2}({\boldsymbol{k}})+|\Delta _{0}|^{2}}<0$ at $k_{y\max }b=\pi /2,3\pi
/2$ and $k_{x\max }=k_{F}$, and two hole pockets with $E_{1}\left( k\right)
=\varepsilon _{+}({\boldsymbol{k}})-\sqrt{\varepsilon _{-}^{2}({\boldsymbol{k%
}})+|\Delta _{0}|^{2}}>0$ at $k_{y\max }b=0,\pi $ and $k_{x\max }=k_{F}\pm
2t_{b}/v_{F}$ (see Fig. \ref{FigPockets}). The hole pockets of the new FS
are the elongated ellipses, satisfying $E_{1}\left( \mathbf{k}\right) =0$
and having the main axes along the vectors $\mathbf{k}_{x}$ and $\mathbf{k}%
_{y}$. Two electron pockets are the similar ellipses, rotated in the $k_{x}$-%
$k_{y}$ plane by the angles
\begin{equation}
\phi _{e}=\pm \arctan \left( 2t_{b}b/\hbar v_{F}\right) .  \label{phie}
\end{equation}

Near the points $\boldsymbol{k}=\boldsymbol{k}_{\max }$, where $\left\vert
\varepsilon _{+}({\boldsymbol{k}})\right\vert $ has a maximum and the small
ungapped pockets get formed, the dispersion (\ref{E1}) rewrites as ($\Delta
k_{y}=k_{y}-k_{y\max }$)
\begin{equation}
E_{1}\left( \Delta k_{y},\varepsilon _{-}\right) \approx -\delta
+a_{1}\left( \Delta k_{y}\right) ^{2}+b_{1}\varepsilon _{-}^{2},
\label{Pocket1}
\end{equation}%
where using (\ref{dispersion}) one obtains%
\begin{equation}
\varepsilon _{-}(\mathbf{k})\approx v_{F}(k_{x}-k_{F})-2t_{b}b\sin (k_{y\max
}b)\Delta \mathbf{k}_{y}/\hbar ,  \label{em}
\end{equation}%
\begin{eqnarray}
\delta &\equiv &\left\vert \varepsilon _{+}({\boldsymbol{k}})\right\vert
_{\max }-|\Delta _{0}|=2t{_{b}^{\prime }-}|\Delta _{0}|,  \notag \\
a_{1} &\approx &4t{_{b}^{\prime }}{b}^{2}\text{ and }b_{1}\approx 1/2\Delta
_{0}.  \label{a1}
\end{eqnarray}%
Here $\delta $ has the meaning of the Fermi energy in these small pockets,%
\cite{Cdelta} and the last term in Eq. (\ref{em}) rotates the electron FS
pockets by the angle (\ref{phie}).

Without DW ordering, the density of states (DoS) of electrons with quasi-1D
dispersion (\ref{1}) is%
\begin{equation*}
\rho _{0}\left( E_{F}\right) =\int \frac{dk_{x}dk_{y}}{\left( 2\pi \right)
^{2}}\delta \left[ \hbar v_{F}\left( k_{x}\pm k_{F}\right) \right] =\frac{1}{%
\pi \hbar v_{F}b}.
\end{equation*}%
Let us estimate the DoS, on the Fermi level in the DW phase when the open
pockets just appear. By definition\cite{AGD}
\begin{equation}
\rho \left( \varepsilon \right) =-\left( 1/\pi \right) \text{Im}\left[ \text{%
Tr}G_{ret}\left( \varepsilon \right) \right] ,  \label{DoSDef}
\end{equation}%
The retarded Green function is obtained from (\ref{GRR}) by the analytical
continuation $i\omega \rightarrow \varepsilon +i0$. Its substitution to (\ref%
{DoSDef}) gives the DoS on the Fermi level (at $\varepsilon =0$)
\begin{equation}
\rho \left( E_{F}\right) =\sum_{\mathbf{k}}\left( \frac{\varepsilon (\mathbf{%
k})}{E_{1}\left( \mathbf{k}\right) }\delta \left[ E_{2}\left( \mathbf{k}%
\right) \right] +\frac{\varepsilon (\mathbf{k})}{E_{2}\left( \mathbf{k}%
\right) }\delta \left[ E_{1}\left( \mathbf{k}\right) \right] \right) ,
\end{equation}%
where $\delta \left[ x\right] $ is the Dirac $\delta $-function. For small
FS pockets, i.e. at $\delta \ll \Delta _{0}$, the residues of the Green
function poles
\begin{equation*}
\frac{\varepsilon (\mathbf{k})}{E_{1}\left( \mathbf{k}\right) }=\frac{%
\varepsilon (\mathbf{k})}{E_{2}\left( \mathbf{k}\right) }\approx \frac{1}{2},
\end{equation*}%
and the contribution of one ungapped FS pocket per one spin orientation is
given by
\begin{eqnarray*}
\rho _{1} &=&\int \frac{dk_{x}d\Delta k_{y}}{\left( 2\pi \right) ^{2}}\frac{%
\delta \left[ a_{1}\left( \Delta k_{y}\right) ^{2}+b_{1}\varepsilon
_{-}^{2}-\delta \right] }{2} \\
&=&\int \frac{dxdy/\left( 2\pi \right) ^{2}}{\hbar v_{F}\sqrt{a_{1}b_{1}}}%
\frac{\delta \left( x^{2}+y^{2}-\delta \right) }{2} \\
&=&\int \frac{dr^{2}/8\pi }{\hbar v_{F}b}\delta \left( r^{2}-\delta \right) =%
\frac{1}{8\pi \hbar v_{F}b}.
\end{eqnarray*}%
There are 8 ungapped pockets for the dispersion (\ref{dispersion}), and the
total DoS on the Fermi level per one spin component in DW state is the same
as without DW ordering:
\begin{equation}
\rho \left( E_{F}\right) =8\rho _{1}=\frac{1}{\pi \hbar v_{F}b}.
\label{DoSDW}
\end{equation}%
This result differs from that in the previously studied models\cite%
{Bilbro,Machida}, where the electron spectrum on the ungapped parts of the
FS does not change after the formation of DW, and the DoS on the Fermi level
reduces when the DW is formed, so that the SC transition temperature reduces
exponentially. In our model, the DoS on the Fermi level in the DW state with
open FS pockets is the same as without DW. Therefore, the SC transition
temperature in our model does not change so strong when the DW ordering with
ungapped pockets is destroyed completely with the restoration of the
metallic state (see Sec. III A for more details). The DoS on the Fermi level
also determines many other physical properties.

\subsection{Stability with respect to superconductivity in the metallic state%
}

The phonon-mediated electron pairing produces only the charge coupling $%
U_{c}\left( \boldsymbol{Q}\right) $, and in the study of superconductivity,
one may use the hamiltonian (\ref{Hint0})-(\ref{vQ1}), neglecting the
coupling $U_{s}\left( \boldsymbol{Q}\right) $ in (\ref{vQ}). Then, in terms
of left and right moving electrons, the interaction Hamiltonian has the form%
\begin{align}
\hat{H}_{\mathrm{int}}& =\frac{1}{2}\sum_{{\boldsymbol{k}}{\boldsymbol{k}}%
^{\prime }{\boldsymbol{Q}}}U_{c}^{b}a_{\alpha }^{\dag R}({\boldsymbol{k}}+{%
\boldsymbol{Q}})a_{\alpha }^{L}({\boldsymbol{k}})a_{\beta }^{\dag L}({%
\boldsymbol{k}}^{\prime }-{\boldsymbol{Q}})a_{\beta }^{R}({\boldsymbol{k}}%
^{\prime })  \notag \\
& +\frac{1}{2}\sum_{{\boldsymbol{k}}{\boldsymbol{k}}^{\prime }{\boldsymbol{Q}%
}}U_{c}^{f}a_{\alpha }^{\dag R}({\boldsymbol{k}}+{\boldsymbol{Q}})a_{\alpha
}^{R}({\boldsymbol{k}})a_{\beta }^{\dag L}({\boldsymbol{k}}^{\prime }-{%
\boldsymbol{Q}})a_{\beta }^{L}({\boldsymbol{k}}^{\prime }).  \label{Hint}
\end{align}%
With two FS sheets in (\ref{1}), it is useful to describe SC in terms of two
Gor'kov functions \cite{AGD}
\begin{equation}
F^{L(R)\,R(L)}(X_{1},X_{2})=<T(\hat{\Psi}^{L(R)}(X_{1})\hat{\Psi}%
^{R(L)}(X_{2}))>,  \label{GF}
\end{equation}%
where $X=(\tau ,\boldsymbol{r})$, and $\hat{\Psi}^{L(R)}(X)$ are the field
operators for the left and right parts of the Brillouin zone, formally,
comprising the electrons with momenta $P_{\parallel }<0$ (L) and $%
P_{\parallel }>0$ (R). The averages in (\ref{GF}) at $\tau _{1}=\tau _{2}+0$
\begin{equation}
\begin{split}
\hat{f}_{\alpha \beta }^{LR}(\mathbf{r})& =<\hat{\Psi}_{\alpha }^{L}(\mathbf{%
r})\hat{\Psi}_{\beta }^{R}(\mathbf{r})>; \\
\hat{f}_{\alpha \beta }^{RL}(\mathbf{r})& =<\hat{\Psi}_{\alpha }^{R}(\mathbf{%
r})\hat{\Psi}_{\beta }^{L}(\mathbf{r})>
\end{split}
\label{GF1t}
\end{equation}%
have the meaning of the Cooper pair wave function and determine the SC order
parameter $\hat{\Delta}_{SC}(\mathbf{r})$. The "hat" above the functions $%
\hat{f}^{LR}(\mathbf{r})$ and $\hat{\Delta}_{SC}(\mathbf{r})$ means that
these functions are operators in the spin space. In the materials with
spatial inversion symmetry, like (TMTSF)$_{2}$PF$_{6}$, one has $\hat{f}%
^{LR}=\pm \hat{f}^{RL}$, and the sign ($\pm $) depends on whether the SC
pairing is singlet (+) or triplet (-). Below we assume the uniform SC order
parameter: $\hat{f}_{\alpha \beta }^{LR}(\mathbf{r})=\hat{f}_{\alpha \beta
}^{LR}$. In the momentum representation Eq. (\ref{GF1t}) rewrites
\begin{equation}
\begin{split}
\hat{f}_{\alpha \beta }^{LR}& =\sum_{\boldsymbol{k}}<a_{\alpha }^{L}(\mathbf{%
k})a_{\beta }^{R}(\mathbf{-k})>; \\
\hat{f}_{\alpha \beta }^{RL}& =\sum_{\boldsymbol{k}}<a_{\alpha }^{R}(\mathbf{%
k})a_{\beta }^{L}(\mathbf{-k})>.
\end{split}
\label{GF1}
\end{equation}

\begin{figure}[tb]
\includegraphics[width=0.49\textwidth]{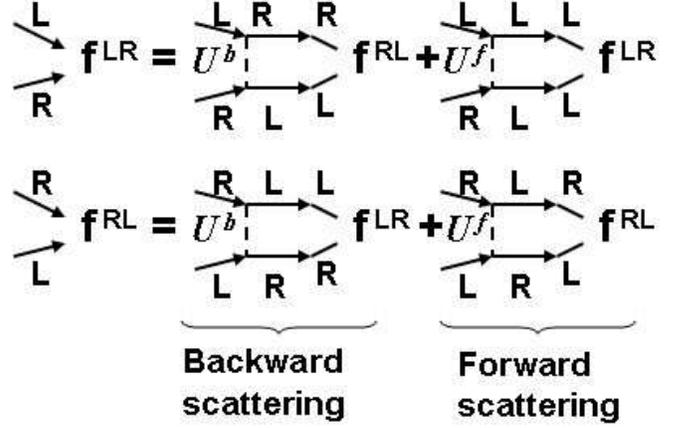}
\caption{The diagram equations for the Gor'kov functions $\hat{f}^{LR}$ and $%
\hat{f}^{RL}$ without DW. The solid lines represent the electron Green
functions: $g^{RR(LL)}\left( \boldsymbol{k},\protect\omega \right) $ in the
metallic state. The dashed lines represent the short-range e--e interaction $%
U_{c}^{f}$ or $U_{c}^{b}$.}
\label{Fig1Eq}
\end{figure}

We introduce the notation for the Cooper bubble:%
\begin{equation}
\Pi _{d}=T\sum_{\mathbf{k},\omega }g^{RR}({\boldsymbol{k}},{\boldsymbol{k}}%
,\omega )g^{LL}(-{\boldsymbol{k}},-{\boldsymbol{k}},-\omega ),  \label{Pd0}
\end{equation}%
where the Green functions $g^{RR\left( LL\right) }({\boldsymbol{k}},{%
\boldsymbol{k}},\omega )$ in the metallic state given by Eq. (\ref{GRR}) at $%
\Delta _{0}=0$. From the Hamiltonian (\ref{Hint}) with definition (\ref{Pd0}%
) one obtains the Gor'kov equations for the onset of SC:%
\begin{equation}
\begin{split}
\hat{f}^{LR}& =-\left( U_{c}^{b}\hat{f}^{RL}+U_{c}^{f}\hat{f}^{LR}\right)
\Pi _{d}, \\
\hat{f}^{RL}& =-\left( U_{c}^{b}\hat{f}^{LR}+U_{c}^{f}\hat{f}^{RL}\right)
\Pi _{d}.
\end{split}
\label{SC0}
\end{equation}%
Eq. (\ref{SC0}) is shown schematically in Fig. \ref{Fig1Eq}. Summation and
subtraction of the two lines in Eq. (\ref{SC0}) gives the equations on SC
transition temperature $T_{c0}^{SC}$ in the metallic state:%
\begin{equation}
\begin{split}
\hat{f}^{LR}+\hat{f}^{RL}& =-\left( U_{c}^{f}+U_{c}^{b}\right) \left( \hat{f}%
^{LR}+\hat{f}^{RL}\right) \Pi _{d}, \\
\hat{f}^{LR}-\hat{f}^{RL}& =-\left( U_{c}^{f}-U_{c}^{b}\right) \left( \hat{f}%
^{LR}-\hat{f}^{RL}\right) \Pi _{d}.
\end{split}
\label{SC0t}
\end{equation}%
The first line in Eq. (\ref{SC0t}) corresponds to singlet, and the second
line to the triplet pairing. Usually, $-U_{c}^{f}-U_{c}^{b}\,>%
\,U_{c}^{b}-U_{c}^{f}$, and the singlet SC transition temperature is higher.
Eq. (\ref{SC0t}) rewrites as
\begin{equation}
1=g\Pi _{d},~g=\max \left\{
-U_{c}^{f}-U_{c}^{b}\,,\,U_{c}^{b}-U_{c}^{f}\right\} .  \label{1gPid}
\end{equation}%
Therefore, in our model one has singlet or triplet superconductivity
depending on the ratio of the coupling constants $U_{c}^{f}$ and $U_{c}^{b}$%
. The nonmagnetic impurities also suppress the triplet SC ordering. The
Cooper bubble $\Pi _{d}$ has the well-known logarithmic singularity,
appearing after the summation over momenta and frequencies:%
\begin{equation}
\Pi _{d}^{met}=\Pi _{d}^{met}\left( T\right) \approx \nu _{F}\ln \left(
\overline{\omega }/T\right) ,  \label{PiMet}
\end{equation}%
where $\overline{\omega }$ is a proper cutoff,\cite{AGD} and $\nu _{F}=\rho
_{0}\left( E_{F}\right) $ is the density of states at the Fermi level. For
the quasi-1D electron spectrum (\ref{1}), $\nu _{F}=1/\pi \hbar v_{F}b$.
From (\ref{1gPid}),(\ref{PiMet}) one obtains the equation for the SC
critical temperature $T_{c0}^{SC}$:
\begin{equation}
1\approx g\nu _{F}\ln (\overline{\omega }/T_{c0}^{SC}).  \label{Tc1}
\end{equation}

\section{SC in the CDW state}

First, we study the SC instability in the CDW state, where the spin
structures of the CDW and SC order parameters do not interfere. As we shall
see below in Sec. IV, the results obtained in this section for the
spin-singlet superconductivity on the CDW background can be applied with
little modification for the triplet superconductivity on the SDW background.
The problem of SC instability and the upper critical field $H_{c2}$ on the
CDW background is important itself. The organic metal $\alpha $-(BEDT-TTF)$%
_{2}$KHg(SCN)$_{4}$ gives an example, where the interplay of
superconductivity and CDW leads to the new SC properties,\cite{CDWSC} and
there are many other CDW superconductors.\cite{Review1}

\begin{figure}[tbh]
\includegraphics[width=0.49\textwidth]{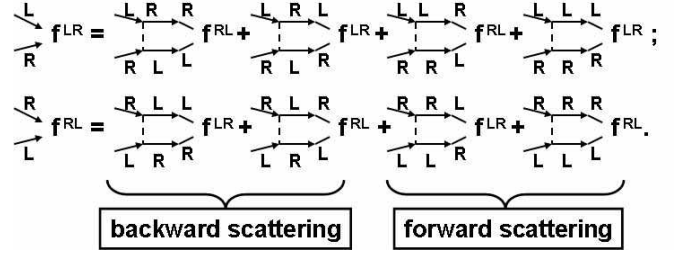}
\caption{The diagram equations for the Gor'kov functions $\hat{f}^{LR}$ and $%
\hat{f}^{RL}$ in the presence of the DW ordering with two coupling
constants. The solid lines represent the electron Green functions $%
g_{_{n}}^{R(L)R(L)}\left( k,\protect\omega \right) $ in the DW state. The
dashed lines represent the short-range e--e interaction, i.e. backward $%
g_{b} $ or forward $g_{f}$ scattering.}
\label{Fig1Eq3}
\end{figure}

The basic equations for the CDW state without superconductivity are obtained
from Eqs. (\ref{hatG})-(\ref{a1}) by removing the spin factor $(\vec{\hat{%
\sigma}}\vec{l})$ from the nondiagonal elements in Eqs. (\ref{GEw})-(\ref%
{hatG}). Thus, the matrix Green function in the uniform CDW state without SC
resembles Eq. (\ref{hatG}):
\begin{equation}
\hat{G}\equiv \left(
\begin{array}{cc}
g^{RR}({\boldsymbol{k}},{\boldsymbol{k}},\omega ) & g^{LR}({\boldsymbol{k}}-{%
\boldsymbol{Q}},{\boldsymbol{k}},\omega ) \\
g^{RL}({\boldsymbol{k}},{\boldsymbol{k}}-{\boldsymbol{Q}},\omega ) & g^{LL}({%
\boldsymbol{k}}-{\boldsymbol{Q}},{\boldsymbol{k}}-{\boldsymbol{Q}},\omega )%
\end{array}%
\right) ,  \label{GCDW}
\end{equation}%
with the matrix components given by Eqs. (\ref{GLR})-(\ref{GRR}). In
addition to the term (\ref{Pd0}), the Cooper bubble on the DW background
contains another term, coming from the nondiagonal elements in the Green
function (\ref{GCDW}):
\begin{equation}
\Pi _{n}=T\sum_{\mathbf{k},\omega }g^{LR}({\boldsymbol{k-Q}},{\boldsymbol{k}}%
,\omega )g^{RL}(-{\boldsymbol{k+Q}},-{\boldsymbol{k}},-\omega ).  \label{Pn}
\end{equation}%
Therefore, the Gor'kov equations on the DW background acquire two additional
terms as compared to Eq. (\ref{SC0}):%
\begin{equation}
\begin{split}
\hat{f}^{LR}& =-\left( U_{c}^{b}\hat{f}^{RL}+U_{c}^{f}\hat{f}^{LR}\right)
\Pi _{d}-\left( U_{c}^{b}\hat{f}^{LR}+U_{c}^{f}\hat{f}^{RL}\right) \Pi _{n},
\\
\hat{f}^{RL}& =-\left( U_{c}^{b}\hat{f}^{LR}+U_{c}^{f}\hat{f}^{RL}\right)
\Pi _{d}-\left( U_{c}^{b}\hat{f}^{RL}+U_{c}^{f}\hat{f}^{LR}\right) \Pi _{n}.
\end{split}
\label{SC0DW}
\end{equation}%
In writing these equations we use that the spin structure of the Gor'kov
functions $\hat{f}^{LR}$\ commutes with the Green functions $g^{R(L)R(L)}({%
\boldsymbol{k}},{\boldsymbol{k}}^{\prime },\omega )$ on the CDW background.
Eq. (\ref{SC0DW}) is shown schematically in Fig. \ref{Fig1Eq3}.

The summation and subtraction of the two lines in Eq. (\ref{SC0DW}) give the
equation on the SC transition temperature for singlet and triplet pairing
respectively:%
\begin{equation}
\begin{split}
\hat{f}^{LR}+\hat{f}^{RL}& =-\left( U_{c}^{f}+U_{c}^{b}\right) \left( \hat{f}%
^{LR}+\hat{f}^{RL}\right) \left( \Pi _{d}+\Pi _{n}\right) , \\
\hat{f}^{LR}-\hat{f}^{RL}& =-\left( U_{c}^{f}-U_{c}^{b}\right) \left( \hat{f}%
^{LR}-\hat{f}^{RL}\right) \left( \Pi _{d}-\Pi _{n}\right) .
\end{split}
\label{SC0tCDW}
\end{equation}%
Below we show that $\Pi _{d}$ and $\Pi _{n}$ have the same sign, and $%
\left\vert \Pi _{d}+\Pi _{n}\right\vert >\left\vert \Pi _{d}-\Pi
_{n}\right\vert $. Therefore, if in the metallic state the transition
temperature to singlet SC $T_{cSC}^{\text{Singlet}}$ is higher than $%
T_{cSC}^{\text{Triplet}}$ to the triplet SC, then on the CDW background $%
T_{cSC}^{\text{Singlet}}>T_{cSC}^{\text{Triplet}}$ is also valid. For SDW,
the interplay of the spin structures of SC and SDW produces important
changes (see Sec. IV).

\subsection{SC instability and transition temperature in the uniform CDW
state}

The equation on the singlet SC transition temperature $T_{cCDW}^{SC}$ on CDW
background, given by the first line in Eq. (\ref{SC0tCDW}), writes down as $%
K_{1}\equiv g\left( \Pi _{d}+\Pi _{n}\right) $ $=1$, where for singlet
pairing $g=-\left( U_{c}^{f}+U_{c}^{b}\right) $. Using $\varepsilon (\mathbf{%
k})=\varepsilon (-{\boldsymbol{k}})$ and substituting (\ref{GLR}),(\ref{GRR}%
), we obtain
\begin{eqnarray}
K_{1} &=&Tg\sum_{\mathbf{k},\omega _{n}}\frac{\omega ^{2}+\left[ \varepsilon
_{-}({\boldsymbol{k}})+\varepsilon _{+}({\boldsymbol{k}})\right]
^{2}+|\Delta _{0}|^{2}}{\left[ \omega ^{2}+E_{1}^{2}({\boldsymbol{k}})\right]
\left[ \omega ^{2}+E_{2}^{2}({\boldsymbol{k}})\right] }  \label{KsCDW} \\
&=&\frac{Tg}{2}\sum_{\mathbf{k},\omega _{n}}\left( \frac{1}{\omega
^{2}+E_{1}^{2}({\boldsymbol{k}})}+\frac{1}{\omega ^{2}+E_{2}^{2}({%
\boldsymbol{k}})}\right)  \label{KCDW1} \\
&=&\frac{g}{v_{F}}\int_{0}^{2\pi /b}\frac{bdk_{y}}{2\pi }\int_{-\overline{%
\omega }}^{\overline{\omega }}\frac{d\varepsilon _{-}}{2\pi }\frac{\tanh %
\left[ E_{1}\left( \mathbf{k}\right) /2T\right] }{E_{1}\left( \mathbf{k}%
\right) }.  \label{Ks1}
\end{eqnarray}%
In writing the second line, Eq. (\ref{KCDW1}), we have substituted (\ref{E1}%
) and used the symmetry of the functions $\varepsilon _{\pm }(k_{y})$: $%
\varepsilon _{+}(k_{y})$ is an even function of $k_{y}$, and $\int
dk_{y}F\left( \varepsilon _{\pm }(k_{y})\right) =0$ for any odd function $%
F\left( \varepsilon \right) $. Let us now rewrite $K_{1}\equiv
K_{ult}+K_{\inf }$, where
\begin{eqnarray}
K_{ult} &\equiv &\frac{g}{v_{F}}\int_{0}^{2\pi /b}\frac{bdk_{y}}{2\pi }%
\int_{|\Delta _{0}|}^{\overline{\omega }}\frac{d\varepsilon _{-}}{\pi }\frac{%
\tanh \left[ E_{1}\left( \mathbf{k}\right) /2T\right] }{E_{1}\left( \mathbf{k%
}\right) }  \notag \\
&\approx &\left[ g/\pi v_{F}\right] \ln \left( \overline{\omega }/\Delta
_{0}\right)  \label{Ko2}
\end{eqnarray}%
contains the ultraviolet logarithmic divergence in expression (\ref{Ks1}),
and%
\begin{equation}
K_{\inf }\equiv \frac{g}{v_{F}}\int_{0}^{2\pi /b}\frac{bdk_{y}}{2\pi }%
\int_{0}^{|\Delta _{0}|}\frac{d\varepsilon _{-}}{\pi }\frac{\tanh \left[
E_{1}\left( \mathbf{k}\right) /2T\right] }{E_{1}\left( \mathbf{k}\right) }
\label{Kinf}
\end{equation}%
may contain the infrared logarithmic divergence if there are electron states
on the Fermi level. At $P>P_{c1}$ the ungapped electron states appear as
small Fermi-surface pockets (see Sec. IIA and Fig. \ref{FigPockets}), or as
the soliton band.\cite{GGPRB2007} In each case, the formed small "Fermi
surface" is subjected to the Cooper instability at rather low temperature,
which signifies the possibility for the onset of SC pairing.

Substituting (\ref{Pocket1}) for $E_{1}({\boldsymbol{k}})$ in Eq. (\ref{Kinf}%
) and introducing $r^{2}\equiv a_{1}\left( \Delta k_{y}\right)
^{2}+b_{1}\varepsilon _{-}^{2}$, we obtain%
\begin{eqnarray}
K_{\inf } &\approx &\frac{N_{P}^{e}gb/v_{F}}{4\pi \sqrt{a_{1}b_{1}}}%
\int_{0}^{\Delta _{0}}\frac{\tanh \left[ \left( \delta -r^{2}\right) /2T%
\right] }{\delta -r^{2}}dr^{2}  \notag \\
&\approx &\frac{N_{P}^{e}g}{2\pi v_{F}\sqrt{2t{_{b}^{\prime }}{/}\Delta _{0}}%
}\ln \left[ C\sqrt{\Delta _{0}\delta }/T\right] ,  \label{K1CDW}
\end{eqnarray}%
where $C\sim 1$ is a numerical constant, and $N_{P}^{e}$ is the number of
ungapped electron pockets on one FS sheet. With the tight-binding dispersion
(\ref{dispersion}), at $2t_{b}^{\prime }>\Delta _{0}$ in each Brillouin zone
$N_{P}^{e}=2$ (see Fig. \ref{FigPockets}). Thus, when the small pockets just
appear, i.e. when $0<2t{_{b}^{\prime }/}\Delta _{0}-1\ll 1$,
\begin{equation}
K_{1}\approx \frac{g}{\pi v_{F}}\left[ \ln \left( \overline{\omega }/\Delta
_{0}\right) +\ln \left( C\sqrt{\Delta _{0}\delta }/T\right) \right] .
\label{K1c}
\end{equation}%
Comparing Eqs. (\ref{K1c}) and (\ref{Tc1}) one obtains, that the SC critical
temperature $T_{cCDW}^{SC}$ in the CDW state is related to the SC transition
temperature $T_{c0}^{SC}$ without CDW as
\begin{equation}
T_{cCDW}^{SC}\approx CT_{c0}^{SC}\sqrt{\delta /\Delta _{0}}\,.  \label{TCDW}
\end{equation}%
This result differs from Eqs. 3.5 and 3.7 of Ref. \cite{Machida}, where $%
T_{cCDW}^{SC}$ was exponentially smaller than $T_{c0}^{SC}$. The origin of
this difference was explained in the end of Sec. IIB, where the DoS on the
Fermi level in the DW state with small open pockets and in the metallic
state were shown to be approximately the same. The assumption that the
electron spectrum in the ungapped FS pockets does not change after the
formation of DW, used in Refs. [\onlinecite{Bilbro,Machida}], is not valid,
especially when the ungapped FS pockets are small. For quasi-1D
tight-binding dispersion (\ref{1})-(\ref{dispersion}), the SC transition
temperature on the CDW background with ungapped FS pockets $T_{cCDW}^{SC}$
is only slightly less than $T_{c0}^{SC}$.

\begin{figure}[tbh]
\label{FigTc} \includegraphics[width=0.49\textwidth]{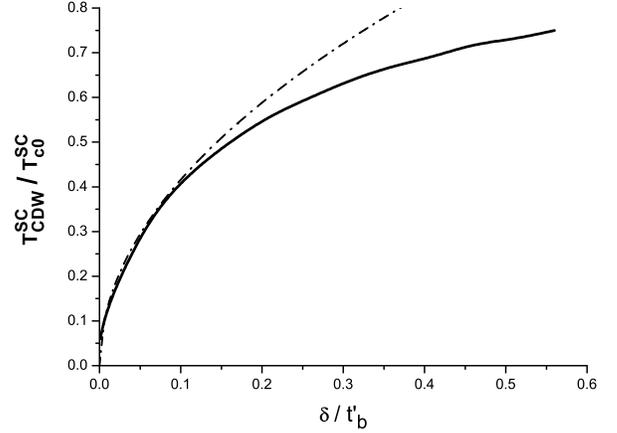}
\caption{The SC transition temperature $T_{cCDW}^{SC}$ on the CDW background
in the pocket scenario. The solid line shows the ratio $%
T_{cCDW}^{SC}/T_{c0}^{SC}$, calculated numerically from Eq. (\protect\ref%
{Ks1}), as function of the size $\protect\delta $ of the ungapped Fermi
surface pockets. The dash-dotted line represents the analytical formula (%
\protect\ref{TCDW}) with the numerically calculated constant $C\approx 1.86$%
. }
\label{FigTSC}
\end{figure}

Eqs. (\ref{Ko2})-(\ref{TCDW}) were derived following Ref. [%
\onlinecite{GGPRB2007}] with logarithmic accuracy, i.e. assuming $\ln \left(
\delta /T\right) \gg 1$ and $\ln \left( \Delta _{0}/\delta \right) \gg 1$.
This accuracy is not sufficient to determine the constant $C$. For more
accurate estimate of the transition temperature $T_{cCDW}^{SC}$, we
calculated the integral (\ref{Ks1}) numerically for the particular
dispersion (\ref{dispersion}). This calculation confirms the approximate
formula (\ref{TCDW}) at $\delta \ll \Delta _{0}$, and gives the value of the
constant $C\approx 1.86$ (see Fig. \ref{FigTSC}). At $\delta /t_{b}^{\prime
}\ll 1$ the analytical and numerical results coincide, while at $\delta \sim
t_{b}^{\prime }$ the ratio $T_{cCDW}^{SC}\left( \delta \right) /T_{c0}^{SC}$
tends to saturate, being always less than unity. In agreement with Eq. (\ref%
{TCDW}), the ratio $T_{cCDW}^{SC}/T_{c0}^{SC}$ is almost independent of $%
T_{c0}^{SC}$.

The above analytical and numerical estimations show, that the CDW state with
ungapped pockets on the Fermi surface is always unstable toward the
formation of the superconductivity, and the SC transition temperature $%
T_{cCDW}^{SC}$ on the CDW background is not very low, being slightly less
than the SC transition temperature $T_{c0}^{SC}$ without CDW. The formation
of the ungapped pockets in the CDW state due to the increase of the
antinesting term at $P>P_{c1}$ is, usually, accompanied by the reduction of
the CDW energy gap $\Delta _{0}$ and, hence, by the fast growth of the size $%
\delta \left( P\right) $ of the ungapped FS pockets. Then, from Eq. (\ref%
{TCDW}) we obtain, that the SC critical temperature $T_{cCDW}^{SC}$ (\ref%
{TCDW}) also grows very rapidly at $P>P_{c1}$. In experiment, this fast
growth of $T_{cCDW}^{SC}\left( P\right) $ above $P_{c1}$ may be similar to
the jump of $T_{cCDW}^{SC}$ from zero to some finite value.

The fluctuations and change of the phonon modes, accompanying the transition
from CDW to metallic state in the whole pressure interval $P_{c1}\lesssim
P\lesssim P_{c1}$, may considerably increase the SC transition temperature $%
T_{cCDW}^{SC}$ and influence the dependence $T_{cCDW}^{SC}\left( P\right) $.
CDW also affects the screening of the Coulomb interaction, which changes the
e--e coupling and the SC transition temperature. Even a small change in the
e--e coupling constant leads to the dramatic changes in the SC transition
temperature.\cite{AGD} An accurate calculation of $T_{cCDW}^{SC}\left(
P\right) $ on the DW background must take these two effects into account,
being beyond the scope of the present paper.

\subsection{Upper critical field $H_{c2}$ in the SC state on the uniform CDW
background}

Upper critical field in superconductors with intrinsic DW ordering was
considered theoretically in the model,\cite{GabovichH} where the DW gap
appears only on those FS sections, where the nesting condition is fulfilled,
while on the rest of FS the electron spectrum has no singularities. This is
not the case for our model (see Sec. II), where the dispersion of the
ungapped electrons in DW state have singularity at $P\rightarrow P_{c1}$,
which affect the SC properties and $H_{c2}$. Other calculations\cite%
{Ro,MachidaComment} of $H_{c2}$ in DW superconductors also use the models
very different from the one considered in Sec. II.

To calculate the upper critical field $H_{c2}$, let us use the
Ginzburg-Landau (G-L) approximation. The main contribution to the gradient
term comes from the ungapped pockets of the Fermi surface ($\left\vert
\varepsilon _{+}({\boldsymbol{k}})\right\vert >\Delta _{0}$). For arbitrary
electron dispersion, the G-L functional was derived in Ref. [%
\onlinecite{Melik}]. The order parameter in the general form is a function
of two wave vectors:%
\begin{equation*}
\Delta \left( \mathbf{k}_{1},\mathbf{k}_{2}\right) =\Delta _{\mathbf{q}%
}\left( \mathbf{k}\right) =\Delta \left( \mathbf{q}\right) \psi \left(
\mathbf{k}\right) ,
\end{equation*}%
where the vector $\mathbf{q=k}_{1}\mathbf{+k}_{2}$ gives the spatial
modulation of the SC order parameter $\Delta \left( \mathbf{r}\right) =\int d%
\mathbf{q}\Delta \left( \mathbf{q}\right) e^{i\mathbf{qr}}$, and $\mathbf{k=k%
}_{1}$ is the momentum of an electron in a Cooper pair. In the case of
s-pairing, $\psi \left( \mathbf{k}\right) =const=1$. For triplet-pairing $%
\Delta _{\mathbf{q}}\left( \mathbf{k}\right) =-\Delta _{\mathbf{q}}\left( -%
\mathbf{k}\right) $, and for the quasi-1D dispersion (\ref{dispersion}) with
two separated FS sheets we may take $\psi \left( \mathbf{k}\right)
=sign\left( k_{x}\right) $. To calculate the gradient term in the
Ginzburg-Landau expansion, we use Eq. (11) of Ref. [\onlinecite{Melik}],
which gives the G-L equation for the order parameter in the form%
\begin{gather}
\sum_{j,k}\frac{1}{2m_{jk}}\left[ \nabla _{j}+2ieA_{j}\left( \mathbf{r}%
\right) \right] \left[ \nabla _{k}+2ieA_{k}\left( \mathbf{r}\right) \right]
\Delta \left( \mathbf{r}\right)  \notag \\
+T\left\{ \frac{T_{c}^{SC}-T}{T_{c}^{SC}}-f\left\vert \Delta \left( \mathbf{r%
}\right) \right\vert ^{2}\Delta \left( \mathbf{r}\right) \right\} =0,
\label{GLEq1}
\end{gather}%
where $A_{j}\left( \mathbf{r}\right) $ is the vector potential,
\begin{eqnarray}
\frac{1}{m_{jk}} &=&\frac{7\xi \left( 3\right) }{12\pi ^{2}T}\int \mathbf{v}%
_{j}\mathbf{v}_{k}\psi ^{2}\left( \mathbf{k}\right) \frac{d\sigma _{F}}{%
\left\vert \mathbf{v}\right\vert }/\int \frac{d\sigma _{F}}{\left\vert
\mathbf{v}\right\vert }  \label{mjk0} \\
&\equiv &\frac{7\xi \left( 3\right) }{12\pi ^{2}T}\frac{\int \mathbf{v}_{j}%
\mathbf{v}_{k}\psi ^{2}\left( \mathbf{k}\right) d^{3}\mathbf{k\,}\delta
\left( E\left( \mathbf{k}\right) -E_{F}\right) }{\int \psi ^{2}\left(
\mathbf{k}\right) d^{3}\mathbf{k\,}\delta \left( E\left( \mathbf{k}\right)
-E_{F}\right) },  \notag
\end{eqnarray}%
\begin{equation}
f=\frac{7\xi \left( 3\right) }{8\left( \pi T\right) ^{2}}\int \psi
^{4}\left( \mathbf{k}\right) \frac{d\sigma _{F}}{\left\vert v_{F}^{\ast
}\right\vert },  \label{fGL}
\end{equation}%
and $\int d\sigma _{F}$ is the integral over the Fermi surface in the
momentum space. Since the G-L equation (\ref{GLEq1}) was derived at $%
T_{c}^{SC}-T\ll T_{c}^{SC}$, one can replace $T$ by $T_{c}^{SC}$ in Eqs. (%
\ref{mjk0}),(\ref{fGL}). Introducing the notations
\begin{equation*}
k_{x}^{\ast }=k_{x}-k_{F};~k_{y}^{\ast }\equiv \Delta k_{y}\left( 2{b}\sqrt{%
2t{_{b}^{\prime }}\Delta _{0}}/\hbar v_{F}\right) ,
\end{equation*}%
one rewrites the dispersion (\ref{Pocket1}) for the hole pockets of the FS
as
\begin{equation}
E\left( \mathbf{k}^{\ast }\right) =\frac{\left( \hbar v_{F}\right) ^{2}}{%
2\Delta _{0}}\left[ k_{y}^{\ast 2}+k_{x}^{\ast 2}\right] -\delta .
\label{Eh1}
\end{equation}%
The Fermi surface, $E\left( \mathbf{k}^{\ast }\right) =0$, is parameterized
by the angle $\phi $, where $\tan \phi \equiv k_{y}^{\ast }/k_{x}^{\ast }\,$%
. The quasi-particle velocity on the FS is a function of this angle:
\begin{eqnarray*}
v_{x} &=&v_{F}\sqrt{2\delta /\Delta _{0}}\cos \phi ; \\
v_{y} &=&\left( 4{b}/\hbar \right) \sqrt{\delta t{_{b}^{\prime }}}\sin \phi .
\end{eqnarray*}%
Performing the integrations in (\ref{mjk0}), we obtain the contribution from
each hole pocket to the tensor (\ref{mjk0}):
\begin{eqnarray}
\left( \frac{1}{m_{xx}}\right) _{h} &=&\frac{7\xi \left( 3\right) v_{F}^{2}}{%
12\pi ^{2}T_{c}^{SC}}\left( \frac{\delta }{\Delta _{0}}\right) ;~
\label{mijh} \\
\left( \frac{1}{m_{yy}}\right) _{h} &=&\frac{14\xi \left( 3\right)
b^{2}t_{b}^{\prime }\delta }{3\pi ^{2}\hbar ^{2}T_{c}^{SC}};~~\left( \frac{1%
}{m_{xy}}\right) _{h}=0.  \notag
\end{eqnarray}%
This mass tensor is very anisotropic:
\begin{equation}
\left( \frac{m_{yy}}{m_{xx}}\right) _{h}=\frac{\hbar ^{2}v_{F}^{2}}{%
8t_{b}^{\prime }\Delta _{0}b^{2}}\sim \left( \frac{\hbar v_{F}}{%
2bt_{b}^{\prime }}\right) ^{2}\gg 1.  \notag
\end{equation}%
The contribution from the electron pockets can be obtained via the rotation
of tensor (\ref{mijh}) by the angles (\ref{phie}):%
\begin{eqnarray*}
\left( \frac{1}{m_{xx}}\right) _{e} &=&\left( \frac{1}{m_{xx}}\right)
_{h}\cos ^{2}\phi _{e}+\left( \frac{1}{m_{yy}}\right) _{h}\sin ^{2}\phi _{e}
\\
\left( \frac{1}{m_{yy}}\right) _{e} &=&\left( \frac{1}{m_{xx}}\right)
_{h}\sin ^{2}\phi _{e}+\left( \frac{1}{m_{yy}}\right) _{h}\cos ^{2}\phi _{e}.
\end{eqnarray*}%
The nondiagonal elements $\left( 1/m_{xy}\right) _{e}$ vanish after the
summation over two electron pockets, which are rotated in the opposite
directions. The total G-L mass tensor is%
\begin{equation*}
\frac{1}{m_{ij}}=4\left[ \left( \frac{1}{m_{ij}}\right) _{e}+\left( \frac{1}{%
m_{ij}}\right) _{h}\right] ,
\end{equation*}%
and, using $\left\vert \phi _{e}\right\vert =\arctan \left( 2t_{b}b/\hbar
v_{F}\right) $ $\ll 1$, we obtain%
\begin{eqnarray}
\frac{1}{m_{xx}} &\approx &\frac{14\xi \left( 3\right) v_{F}^{2}}{3\pi
^{2}T_{c}^{SC}}\left( \frac{\delta }{\Delta _{0}}\right) ;  \label{mGL} \\
\frac{1}{m_{yy}} &\approx &\frac{7\xi \left( 3\right) v_{F}^{2}}{3\pi
^{2}T_{c}^{SC}}\left( \frac{\delta }{\Delta _{0}}\right) \left( \frac{2t_{b}b%
}{\hbar v_{F}}\right) ^{2}.  \notag
\end{eqnarray}%
From the Ginzburg-Landau equation one obtains the $i$-component of the upper
critical field
\begin{equation}
H_{c2}^{i}=e_{ijk}\frac{\left( T_{c}^{SC}-T\right) c}{e\hbar }\sqrt{%
m_{j}m_{k}},  \label{Hcm}
\end{equation}%
where $e_{ijk}$ is the antisymmetric tensor of rang 3. For $H\parallel z$,
the substitution of (\ref{mGL}) in (\ref{Hcm}) gives%
\begin{equation}
H_{c2}^{z}=C_{1}\cdot \left( \frac{T_{c}^{SC}}{\delta }\right) \frac{c\left(
T_{c}^{SC}-T\right) }{bv_{F}e},  \label{Hc2z}
\end{equation}%
where
\begin{equation}
C_{1}=\frac{3\pi ^{2}}{7\xi \left( 3\right) \sqrt{2}}\left( \frac{\Delta _{0}%
}{2t_{b}}\right) .  \label{Hcz}
\end{equation}%
The estimate of the constant (\ref{Hcz}) is very sensitive to the electron
dispersion (\ref{dispersion}), e.g., to the presence of the fourth harmonic $%
2t_{4}\cos \left( 4k_{y}b\right) $ in Eq. (\ref{dispersion}). The fourth
harmonic with $t_{4}/t_{2}>0$ increases the size $\delta $ of the ungapped
hole pockets at $k_{y}b\approx \pi n$ by $2t_{4}$, reducing by the same
amount the size of the electron pockets at $k_{y}b\approx \pi \left(
n+1/2\right) $. If $2t_{4}>\delta $, the electron pockets disappear, and
only the hole pockets contribute to the mass tensor (\ref{mjk0}). The total
mass tensor is then very anisotropic and given by Eq. (\ref{mijh})
multiplied by the number of the hole pockets. Its substitution to Eq. (\ref%
{Hcm}) gives (we take $\Delta _{0}/2t_{b}^{\prime }\approx 1$)
\begin{equation}
C_{1}=3\pi ^{2}/14\xi \left( 3\right) =1.76,  \label{Hcz1}
\end{equation}%
which is greater than (\ref{Hcz}) by a factor $\sim t_{b}/t_{b}^{\prime }\,$%
. The similar increase of the constant $C_{1}$ also appears if the DW wave
vector $\mathbf{Q}$ shifts from $\mathbf{Q}_{0}=\left( 2k_{F},\pi /b\right) $%
, so that the electron pockets disappear, while the size $\delta $ of the
hole pockets increases. Accurate calculation of the constant $C_{1}$
requires the detailed knowledge of electron dispersion.

According to Eq. (\ref{Hc2z}), $H_{c2}^{z}$ diverges as $P-P_{c1}\rightarrow
0$. If $\delta \propto P-P_{c1}$,\cite{Cdelta} assuming $T_{c}^{SC}\approx
const$, one obtains $H_{c2}^{z}\propto 1/\left( P-P_{c1}\right) $. In our
model, according to (\ref{TCDW}), $T_{c}^{SC}\propto \sqrt{\delta }\propto
\sqrt{P-P_{c1}}$, and from (\ref{Hc2z}) one obtains the square-root
divergence of $H_{c2}^{z}$:
\begin{equation}
H_{c2}^{z}\left( P\right) \approx H_{c0}/\sqrt{P/P_{c1}-1}.  \label{HcP}
\end{equation}%
The divergence of $H_{c2}^{z}$ as pressure approaches $P_{c1}$ has been
observed in the mixed state in (TMTSF)$_{2}$PF$_{6}$ (see Fig. 2 in Ref. [%
\onlinecite{Hc2Pressure}]) and also in $\alpha $-(BEDT-TTF)$_{2}$KHg(SCN)$%
_{4}$ (see Figs. 5 and 6 of Ref. [\onlinecite{CDWSC}]). To explain this
dependence $H_{c2}^{z}\left( P-P_{c1}\right) $ in the scenario of the
macroscopic spatial phase separation,\cite{Hc2Pressure} the width $d_{S}$ of
the superconducting domains must be taken much smaller than the SC coherence
length $\xi _{SC}$, because in a thin type-II-superconductor slab of
thickness $d_{s}\ll \xi _{SC}$ the upper critical field $H_{c2}$ is higher
than in the bulk superconductor by a factor (see Eq. 12.4 of Ref. [%
\onlinecite{Ketterson}])%
\begin{equation}
H_{c2}/H_{c2}^{0}\approx \sqrt{12}\xi _{SC}/d_{s}.  \label{Hslab}
\end{equation}%
In the discussion of Ref. [\onlinecite{Hc2Pressure}] the penetration length $%
\lambda $ instead of the coherence length $\xi _{SC}$ enters the expression
for $H_{c2}\,$\ in a thin superconducting slab, which is correct only for
type I superconductors. If $d_{s}\ll \xi _{SC}$, the domain size $d_{s}$ is
of the order of the DW coherence length, which may cost additional energy
because of the change of the DW structure. Then, the soliton scenario\cite%
{GGPRB2007} of the DW/SC structure is possible.

The available experimental data in (TMTSF)$_{2}$PF$_{6}$ is not sufficient
to determine the behavior of $T_{c}^{SC}\left( P\right) $ close to the
critical pressure $P=P_{c1}$, which is important for quantitative comparison
of the dependence $H_{c2}\left( P\right) $ in Eqs. (\ref{Hc2z}),(\ref{HcP})
with experiment. We only make the order-of-magnitude comparison with
experiment to check if the model is reasonable. In (TMTSF)$_{2}$PF$_{6}$ the
Fermi velocity $v_{F}\approx 2\cdot 10^{7}$ cm/sec, the interchain spacing $%
b\approx 7.7\mathring{A}$, and $\Delta _{0}/2t_{b}\approx t_{b}^{\prime
}/t_{b}\approx 0.1$. Substituting this and $\delta \approx T_{c}^{SC}$ into
Eq. (\ref{Hc2z}) gives the slope $dH_{c2}^{z}/dT\approx $ $1\,$Tesla$%
/^{\circ }K$ in a reasonable agreement with the experimental data at $%
P\rightarrow P_{c1}$ (see Fig. 2 in Ref. [\onlinecite{Hc2Pressure}]).

In $\alpha $-(BEDT-TTF)$_{2}$KHg(SCN)$_{4}$ the Fermi velocity\cite{KHgVf} $%
v_{F}\approx 6.5\cdot 10^{6}$, the lattice constant\cite{KHgLattice} $%
b\approx 10\mathring{A}$,~the SC and CDW transition temperatures\cite{CDWSC}
$T_{c}^{SC}\approx 0.1K$ and $T_{c}^{CDW}\approx 8K$. Although the original
Fermi surface in this compound possesses the quasi-2D pockets in addition to
the quasi-1D sheets subjected to the CDW instability, the quasi-1D FS sheets
seem to play an important role in the formation of SC, because
superconductivity appears in the presence of CDW (at $P<P_{c}$) with
approximately the same transition temperature $T_{c}^{SC}$ as in the absence
of CDW at $P>P_{c}$. Hence, SC and CDW share the same quasi-1D conducting
band. In this compound $P_{c}\approx 2.5\,$kbar,\cite{CDWSC} while $P_{c1}$
and $\delta \left( P\right) $ are not known. Probably, even at ambient
pressure $P>P_{c1}$ and $\delta \lesssim \Delta _{0}$. Substitution of $%
\delta \approx \Delta _{0}/2$ to Eq. (\ref{Hc2z}) gives the estimate $%
dH_{c2}^{z}/dT\approx $ $\left( 1-T/T_{c}^{SC}\right) \cdot 3.2mT$ in
agreement with experiment (see Figs. 5 of Ref. [\onlinecite{CDWSC}]).

\section{Superconducting instability in the SDW state}

The Green functions in the SDW state are given by Eqs. (\ref{hatG})-(\ref%
{GRR}), and the Gor'kov equations for SC in the SDW state, shown
schematically in Fig. \ref{Fig1Eq3}, write down as%
\begin{widetext}
\begin{gather}
\hat{f}^{LR}=-TU_{c}^{b}\sum_{\mathbf{k},\omega }\left[
g^{RR}({\boldsymbol{k }},{\boldsymbol{k}},\omega )
\hat{f}^{RL}g^{LL}(-{\boldsymbol{k}},-{ \boldsymbol{k}},-\omega
)^{T} +  g^{LR}({\boldsymbol{k-Q}},{\boldsymbol{k}},\omega
)(\vec{\hat{\sigma} }\vec{l})
\hat{f}^{LR}(\vec{\hat{\sigma}}\vec{l})^{T}
g^{RL}({\boldsymbol{Q}}-{\boldsymbol{k}},-{\boldsymbol{k}},-\omega )\right]
\notag \\
-TU_{c}^{f}\sum_{\mathbf{k},\omega }\left[
g^{LL}(-{\boldsymbol{k}},-{ \boldsymbol{k}},-\omega
)\hat{f}^{LR}g^{RR}({\boldsymbol{k}},{\boldsymbol{k}} ,\omega
)^{T}+ g^{RL}(-{\boldsymbol{k+Q}},-{\boldsymbol{k}},-\omega
)(\vec{\hat{ \sigma}}\vec{l})
\hat{f}^{RL}(\vec{\hat{\sigma}}\vec{l})^{T} g^{LR}({
\boldsymbol{k-Q}},{\boldsymbol{k}},\omega )\right] \label{fBothR}
\end{gather}
and%
\begin{gather}
\hat{f}^{RL}=-TU_{c}^{b}\sum_{\mathbf{k},\omega }\left[ g^{LL}(-{\boldsymbol{%
k}},-{\boldsymbol{k}},-\omega )\hat{f}^{LR}g^{RR}({\boldsymbol{k}},{%
\boldsymbol{k}},\omega )^{T}+ g^{RL}({\boldsymbol{Q}}-{\boldsymbol{k}},-{\boldsymbol{k}},-\omega )(%
\vec{\hat{\sigma}}\vec{l})\hat{f}^{RL}(\vec{\hat{\sigma}}\vec{l})^{T}g^{LR}({%
\boldsymbol{k-Q}},{\boldsymbol{k}},\omega )\right]  \notag \\
-TU_{c}^{f}\sum_{\mathbf{k},\omega }\left[ g^{RR}({\boldsymbol{k}},{%
\boldsymbol{k}},\omega )\hat{f}^{RL}g^{LL}(-{\boldsymbol{k}},-{\boldsymbol{k}%
},-\omega )^{T}+ g^{LR}({\boldsymbol{k-Q}},{\boldsymbol{k}},\omega )(\vec{\hat{\sigma}%
}\vec{l})\hat{f}^{LR}(\vec{\hat{\sigma}}\vec{l})^{T}g^{RL}({\boldsymbol{Q}}-{%
\boldsymbol{k}},-{\boldsymbol{k}},-\omega )\right] .  \label{fBothL}
\end{gather}%
The spin structure of the Gor'kov functions $\hat{f}^{LR}$, which\ depend on
the type of SC pairing, interferes with the spin structure $(\vec{\hat{\sigma%
}}\vec{l})$ of the SDW order parameter. This considerably
changes the properties of SC on the SDW background as compared
those on CDW background, studied in Sec. III.

With notations (\ref{Pd0}) and (\ref{Pn}), Eqs. (\ref{fBothR}) and (\ref{fBothL}) rewrite as
\begin{align}
\label{Eqs} \hat{f}^{LR}+\hat{f}^{RL}& =-T\left(
U_{c}^{b}+U_{c}^{f}\right) \left[ \Pi _{d}\left( \hat{f}^{RL}+\hat{f}^{LR}\right) +\Pi _{n}(%
\vec{\hat{\sigma}}\vec{l})\left( \hat{f}^{LR}+\hat{f}^{RL}\right) (\vec{\hat{%
\sigma}}\vec{l})^{T}\right]  \\
\label{Eqt}
\hat{f}^{LR}-\hat{f}^{RL}& =-T\left(
U_{c}^{b}-U_{c}^{f}\right) \left[ \Pi _{d}\left( \hat{f}^{RL}-\hat{f}^{LR}\right) +\Pi _{n}(%
\vec{\hat{\sigma}}\vec{l})\left( \hat{f}^{LR}-\hat{f}^{RL}\right) (\vec{\hat{%
\sigma}}\vec{l})^{T}\right] .
\end{align}
\end{widetext}

\subsection{SC transition temperature}

\subsubsection{Singlet pairing.}

For spin-singlet paring $\hat{f}^{LR}=\hat{f}^{LR}=\hat{\sigma}_{y}f^{LR}$.
Since
\begin{equation}
\hat{\sigma}_{y}(\vec{\hat{\sigma}}\vec{l})^{T}=-(\vec{\hat{\sigma}}\vec{l})%
\hat{\sigma}_{y},  \label{sigmas}
\end{equation}%
Eq. (\ref{Eqs}) becomes
\begin{equation}
f^{LR}+f^{RL}=-\left( U_{c}^{b}+U_{c}^{f}\right) \left( f^{LR}+f^{RL}\right)
\left( \Pi _{d}-\Pi _{n}\right) .  \label{fsSDW}
\end{equation}%
Substituting Eqs. (\ref{Pd0}),(\ref{Pn}) to (\ref{Eqs}) and using $%
\varepsilon (\mathbf{k})=\varepsilon (-{\boldsymbol{k}})$ we rewrite
equation (\ref{fsSDW}) as $K_{SDW}^{s}=1$, where
\begin{equation}
K_{SDW}^{s}=Tg\sum_{\mathbf{k},\omega _{n}}\frac{\omega ^{2}+\left[
\varepsilon _{-}({\boldsymbol{k}})+\varepsilon _{+}({\boldsymbol{k}})\right]
^{2}-|\Delta _{0}|^{2}}{\left[ \omega ^{2}+E_{1}^{2}({\boldsymbol{k}})\right]
\left[ \omega ^{2}+E_{2}^{2}({\boldsymbol{k}})\right] }=1.  \label{K1}
\end{equation}%
This formula differs from the similar equation (\ref{KsCDW}) for CDW by the
sign before $|\Delta _{0}|^{2}$ in the numerator. This sign change, coming
from the interplay (\ref{sigmas}) of the spin structures of SDW and SC order
parameters, is crucial for the SC transition temperature. As in the case of
CDW, the ungapped pockets of the FS appear when $\left\vert \varepsilon _{+}(%
{\boldsymbol{k}})\right\vert _{\max }>|\Delta _{0}|$, and these pockets are
responsible for the low-energy logarithmic singularity of $K_{1}$ at $%
T\rightarrow 0$. If the system is close to the phase transition at $P=P_{c1}$%
, where these pockets just appear, the antinesting term in the electron
dispersion only slightly exceeds the SDW gap, and $\left\vert \varepsilon
_{+}({\boldsymbol{k}})\right\vert _{\max }-|\Delta _{0}|=\delta \ll |\Delta
_{0}|$. Then in these ungapped pockets $\left\vert \varepsilon _{-}({%
\boldsymbol{k}})\right\vert \sim \delta \ll |\Delta _{0}|$, and the
numerator in (\ref{K1}) near $\omega \rightarrow 0$ has the smallness $%
\delta /|\Delta _{0}|\ll 1$ as compared to the case of CDW. This leads to\
the same smallness of the logarithmically singular term in $K_{SDW}^{s}$ at $%
T\rightarrow 0$. Instead of the Eq. (\ref{Ks1}) one now obtains
\begin{equation}
K_{SDW}^{s}\approx \frac{g}{2}\sum_{\mathbf{k}}\frac{\tanh \left[
E_{1}\left( \mathbf{k}\right) /2T\right] }{E_{1}\left( \mathbf{k}\right) }%
\left( 1-\frac{4|\Delta _{0}|^{2}}{E_{2}^{2}\left( \mathbf{k}\right) }%
\right) .  \label{K1SDW}
\end{equation}%
When the ungapped pockets are small, the extra factor $\left[ 1-4|\Delta
_{0}|^{2}/E_{2}^{2}\left( \mathbf{k}\right) \right] \sim \delta /\Delta
_{0}\ll 1$ makes the infrared divergent term in expression (\ref{K1SDW})
much smaller than in Eq. (\ref{Ks1}) for the CDW background. Therefore, the
spin-singlet SC transition temperature on the SDW background is
exponentially smaller as compared to Eq. (\ref{TCDW}):%
\begin{equation}
T_{cSDW}^{SC}\sim \,\sqrt{\Delta _{0}\delta }\left( T_{c0}^{SC}/\Delta
_{0}\right) ^{\left( \Delta _{0}/\delta \right) }.  \label{TSDW}
\end{equation}%
The estimates (\ref{TCDW}),(\ref{TSDW}) depend strongly on the electron
dispersion.

\subsubsection{Triplet pairing.}

The smallness $\sim \delta /\Delta _{0}$ of the numerator in Eq. (\ref{K1})\
always appears for spin-singlet SC pairing on the SDW background. However,
it does not necessarily appear for spin-triplet paring. The triplet order
parameter has the spin structure $\hat{f}^{LR}=\hat{\sigma}_{y}\left(
\mathbf{\hat{\sigma}\vec{d}}\right) f^{LR}$. Substituting it together with $%
f^{RL}=-f^{LR}$\ into (\ref{Eqt}), using $(\vec{\hat{\sigma}}\vec{l})\left(
\mathbf{\hat{\sigma}\vec{d}}\right) =-\left( \mathbf{\hat{\sigma}\vec{d}}%
\right) (\vec{\hat{\sigma}}\vec{l})+2\left( \mathbf{\vec{d}}\vec{l}\right) $
and
\begin{equation}
(\vec{\hat{\sigma}}\vec{l})\left( \mathbf{\hat{\sigma}\vec{d}}\right) \hat{%
\sigma}_{y}(\vec{\hat{\sigma}}\vec{l})^{T}=\left( \mathbf{\hat{\sigma}\vec{d}%
}\right) \hat{\sigma}_{y}-2\left( \mathbf{\vec{d}}\vec{l}\right) (\vec{\hat{%
\sigma}}\vec{l})\hat{\sigma}_{y},  \label{sigmat}
\end{equation}%
we obtain the self-consistency equation%
\begin{gather}
\left( f^{LR}-f^{RL}\right) \left( \mathbf{\hat{\sigma}\vec{d}}\right)
=\left( U_{c}^{b}-U_{c}^{f}\right) \left( f^{LR}-f^{RL}\right)  \label{ftSDW}
\\
\times \left( \left( \mathbf{\hat{\sigma}\vec{d}}\right) \Pi _{d}-\left[
\left( \mathbf{\hat{\sigma}\vec{d}}\right) -2\left( \mathbf{\vec{d}}\vec{l}%
\right) (\vec{\hat{\sigma}}\vec{l})\right] \Pi _{n}\right) .  \notag
\end{gather}%
For $\mathbf{\vec{d}\parallel }\vec{l}$ the equation on SC transition
temperature is the same as in the case of singlet SC on the CDW background
[see the first line of Eq. (\ref{SC0tCDW})] with only the change of the
coupling constant from $U_{c}^{b}+U_{c}^{f}$ to $U_{c}^{b}-U_{c}^{f}$.
Hence, at $U_{c}^{f}\ll U_{c}^{b}$, the SC transition temperature $%
T_{c}^{SC} $ for $\mathbf{\vec{d}\parallel }\vec{l}$ is approximately given
by Eq. (\ref{TCDW}). For $\mathbf{\vec{d}\perp }\vec{l}$ one obtains the
smallness $\sim \delta /\Delta _{0}$ of the infrared-divergent term in the
Cooper bubble, as in the case of spin-singlet SC on the SDW background. Then
the SC transition temperature $T_{c}^{SC}$ is roughly given by Eq. (\ref%
{TSDW}), being exponentially smaller than in the case $\mathbf{\vec{d}%
\parallel }\vec{l}$. For other mutual orientation of vectors $\mathbf{\vec{d}%
}$ and $\vec{l}$, the spin structures of left and right parts of Eq. (\ref%
{ftSDW}) do not coincide, which means the possible mixing of singlet and
triplet states.

\subsection{Upper critical field}

As we have shown above, the spin-singlet superconductivity, appearing on the
SDW background, has vanishing critical temperature. Hence, we consider only
the triplet superconductivity at $\mathbf{\vec{d}\parallel }\vec{l}$, which
corresponds to the highest critical temperature. For the triplet
superconductivity, the paramagnetic spin effect of magnetic field due to the
interaction with electron spin does not lead to the suppression of
superconductivity, and the upper critical field $H_{c2}$ is completely
determined by the orbital electron motion. In the scenario of ungapped
pockets, the upper critical field $H_{c2}\ $on SDW background at $\mathbf{%
\vec{d}\parallel }\vec{l}$ $\ $is approximately the same as for SC on the
CDW background and is given by Eq. (\ref{Hc2z}).

\section{Summary}

We investigated the structure and the properties of superconductivity,
appearing on the uniform DW background and sharing with DW the common
conducting band. The onset of superconductivity requires ungapped electron
states on the Fermi level, which appear at pressure $P>P_{c1}$, i.e. when
the nesting of the FS is spoiled. There are two possible microscopic
structures of the background DW state with such ungapped states on the Fermi
level: (a) the DW energy gap does not cover the whole Fermi surface, i.e.
there are\ ungapped FS pockets, and (b) the DW\ order parameter is not
spatially uniform, and the soliton band get formed.\cite{GGPRB2007} In this
paper the first scenario is considered in detail. The approach of Ref. \cite%
{GGPRB2007} is generalized to the more realistic e-e interaction, which
includes two coupling constants. It is shown, that the electron dispersion
in the ungapped FS pockets on the DW background is strongly different from
that in the metallic state, so that even very small ungapped FS pockets
create rather high DoS on the Fermi level. This fact makes our results very
dissimilar to many previous theoretical approaches, where the electron
dispersion on the unnested parts of FS in DW state was taken the same as in
the metallic state.\cite{Bilbro,Machida,Psaltakis,GabovichH,Ro,MachidaH} For
the tight-binding dispersion (\ref{1}),(\ref{dispersion}) the DoS on the
Fermi level in the DW state with small ungapped FS pockets is the same as in
the metallic state without DW [see Eq. (\ref{DoSDW})]. Therefore, the SC
transition temperature $T_{cDW}^{SC}$ on the DW background with such
ungapped FS pockets (i.e., at pressure $P>P_{c1}$) is not exponentially
smaller than the SC transition temperature $T_{c0}^{SC}$ in the metallic
state [see Eq. (\ref{TCDW})], and the quantum critical fluctuations at $%
P\approx P_{c1}$ may increase $T_{cDW}^{SC}$ to the value even higher than $%
T_{c0}^{SC}$.

The DW background considerably changes the SC properties. The upper critical
field $H_{c2}$ has unusual pressure dependence [see Eq. (\ref{HcP})] and may
considerably exceed $H_{c2}$ without DW background. According to Eqs. (\ref%
{Hc2z}) and (\ref{HcP}), $H_{c2}$ even diverges as $P\rightarrow P_{c1}$;
this divergence is cut off at $\delta \sim T_{c}^{SC}$. The SDW background
strongly suppresses the spin-singlet superconductivity, while the triplet SC
with certain spin polarization ($\mathbf{\vec{d}\parallel }\vec{l}$) on the
SDW background behaves similarly to the singlet SC on the CDW background.
This means that the SDW background spares the formation of triplet
superconductivity compared to the spin-singlet SC. If both types of SC are
possible, the system with SDW background will choose the triplet SC, even if
it would choose singlet SC without SDW background. The results obtained are
in good agreement with experimental observations in two organic metals
(TMTSF)$_{2}$PF$_{6}$ and $\alpha $-(BEDT-TTF)$_{2}$KHg(SCN)$_{4}$, where SC
coexists with SDW and CDW states respectively, giving an alternative to Ref.
[\onlinecite{Hc2Pressure}] explanation of the unusual pressure dependence of
$H_{c2}$ in (TMTSF)$_{2}$PF$_{6}$ and some other compounds.

\section{Acknowledgement}

The work was supported by RFBR No 06-02-16551 and 06-02-16223, and by
MK-4105.2007.2.


\begin{thebibliography}{99}
\bibitem{Review1} A.M. Gabovich, A.I. Voitenko, J.F. Annett and M. Ausloos,
Supercond. Sci. Technol. 14, R1-R27 (2001).

\bibitem{Levin} K. Levin, D. L. Mills, and S. L. Cunningham, Phys. Rev. B
\textbf{10}, 3821 (1974).

\bibitem{Balseiro} C. A. Balseiro and L. M. Falicov, Phys. Rev. B \textbf{20}%
, 4457 (1979).

\bibitem{Milans} L. Milans del Bosch and Felix Yndurain, Phys. Rev. B
\textbf{41}, 2540 (1990).

\bibitem{Bilbro} Griff Bilbro and W. L. McMillan, Phys. Rev. B \textbf{14},
1887 (1976).

\bibitem{Machida} K. Machida, J. Phys. Soc. Jpn. \textbf{50}, 2195 (1981).

\bibitem{Psaltakis} G. C Psaltakis, J. Phys. C: Solid State Phys. \textbf{17}%
, 2145 (1984).

\bibitem{Vuletic} T. Vuletic, P. Auban-Senzier, C. Pasquier et al., Eur.
Phys. J. B \textbf{25}, 319 (2002).

\bibitem{CDWSC} D. Andres, M. V. Kartsovnik, W. Biberacher, K. Neumaier, E.
Schuberth, and H. M\"{u}ller, Phys. Rev. B \textbf{72}, 174513 (2005).

\bibitem{Clogston} A. M. Clogston, Phys. Rev. Lett. \textbf{9}, 266 (1962);
B.S. Chandrasekhar, Appl. Phys. Lett. 1, 7 (1962).

\bibitem{LeeTripletMany} I. J. Lee, M. J. Naughton, G. M. Danner, and P. M.
Chaikin, Phys. Rev. Lett. \textbf{78}, 3555 (1997); I. J. Lee, P. M.
Chaikin, and M. J. Naughton, Phys. Rev. B \textbf{62}, R14 669 (2000);

\bibitem{LeeAngularHc} I. J. Lee, P. M. Chaikin, and M. J. Naughton, Phys.
Rev. B \textbf{65}, 180502(R) (2002).

\bibitem{LeeKnightShift} I. J. Lee, S. E. Brown, W. G. Clark, M. J. Strouse,
M. J. Naughton, W. Kang, and P. M. Chaikin, Phys. Rev. Lett. \textbf{88},
017004 (2001); I.J. Lee, D. S. Chow, W. G. Clark, M. J. Strouse, M. J.
Naughton, P. M. Chaikin, and S. E. Brown, Phys. Rev. B \textbf{68}, 092510
(2003).

\bibitem{Hc2Pressure} I. J. Lee, P. M. Chaikin, and M. J. Naughton, Phys.
Rev. Lett. \textbf{88}, 207002 (2002).

\bibitem{LeeBrown} I. J. Lee, S. E. Brown, W. Yu, M. J. Naughton, and P. M.
Chaikin, Phys. Rev. Lett. \textbf{94}, 197001 (2005).

\bibitem{GG} L.P. Gor'kov, P.D. Grigorev, Europhys. Lett. \textbf{71}, 425
(2005).

\bibitem{GGPRB2007} L.P. Gor'kov, P.D. Grigoriev, Phys. Rev. B \textbf{75},
020507(R) (2007).

\bibitem{BrazKirovaReview} S.A. Brazovskii and N.N. Kirova, Sov. Sci. Rev. A
Phys., \textbf{5}, 99 (1984).

\bibitem{SuReview} W.P. Su and J. R. Schrieffer, \emph{Physics in One
Dimension}/ Ed. by J. Bernascony and T. Schneider, Springer series in Solid
State Sciences, Berlin, Heidelberg and New York: Springer, 1981.

\bibitem{Brown} Stuart Brown, unpublished.

\bibitem{BCS} J. Bardeen, L. N. Cooper, and J. R. Schrieffer, Phys. Rev.
\textbf{108}, 1175 (1957).

\bibitem{CommentStrongH} The generalization of the mean-field description of
DW in quasi-1D metals in the presence of magnetic field was elaborated in
Ref. \cite{BZ} for SDW and in Refs. [\onlinecite{ZBM,GLCDW}] for CDW. Strong
magnetic field, acting on CDW state, mixes the SDW and CDW order parameters%
\cite{ZBM,GLCDW} and may lead to the series of phase transitions between the
states with different quantized values of the nesting vector.\cite{OrbQuant}
Strong magnetic field, acting on metals with imperfect nesting, can lead to
the field-induced DW.\cite{FIDW}

\bibitem{BZ} A. Bjeli\v{s} and D. Zanchi, Phys. Rev. B \textbf{49}, 5968
(1994).

\bibitem{ZBM} D.~Zanchi, A.~Bjeli\v{s}, and G.~Montambaux, Phys. Rev. B
\textbf{53}, 1240 (1996).

\bibitem{GLCDW} P.D. Grigoriev and D.S. Lyubshin, Phys. Rev. B \textbf{72},
195106 (2005).

\bibitem{OrbQuant} D. Andres, M. V. Kartsovnik, P. D. Grigoriev, W.
Biberacher, and H. M\"{u}ller, Phys. Rev. B \textbf{68}, 201101(R) (2003).

\bibitem{FIDW} L.P. Gor'kov and A.G. Lebed, J. Phys. (Paris) 45, L433
(1984); G. Montambaux, M. Heritier, and P. Lederer, Phys. Rev. Lett. 55,
2078 (1985); A.G. Lebed, Phys. Rev. Lett. 88, 177001 (2002); A.G. Lebed,
JETP Lett. 72, 141 (2000) [Pis'ma Zh. Teor. Eksp. Fiz. 72, 205 (2000)].

\bibitem{Cdelta} Usually, $\delta \propto P-P_{c1}$. However, the reduction
of the energy gap $\Delta _{0}$, accompaning the formation of ungapped
pockets and the reduction of the nested part of FS, can make the growth $%
\delta (P)$ faster at $P=P_{c1}$.

\bibitem{AGD} A.A. Abrikosov, L.P. Gor'kov and I.E. Dzyaloshinskii,
\textquotedblright Methods of quantum field theory in statistical
physics\textquotedblright , Dover Publications, INC., New York 1977.

\bibitem{GabovichH} A. M. Gabovich and A. S. Shpigel, Phys. Rev. B \textbf{38%
}, 297 (1988).

\bibitem{Ro} Charles Ro and K. Levin, Phys. Rev. B 29, 6155 (1984).

\bibitem{MachidaComment} Proposed in Ref. [\onlinecite{MachidaH}] appearence
of an additional spatially modulated SC order parameter $\Delta _{SC}\left(
Q_{N}\right) $ seems to be negligible (or wrong) when the Fermi energy is
much greater than the DW and SC order parameters.

\bibitem{MachidaH} Kazushige Machida, Tamotsu K\={o}nyama, and Takeo
Matsubara, Phys. Rev. B \textbf{23}, 99 (1981).

\bibitem{Melik} L.P. Gor'kov and T.K. Melik-Barkhudarov, JETP \textbf{18},
1031 (1963) [J. Exp. Teor. Fiz. \textbf{45}, 1493 (1963)].

\bibitem{Ketterson} J. B. Ketterson and S. N. Song, \textit{Superconductivity%
}, Cambridge University Press, 1999.

\bibitem{KHgVf} A. E. Kovalev, S. Hill, and J. S. Qualls, Phys. Rev. B
\textbf{66}, 134513 (2002).

\bibitem{KHgLattice} Ryusuke Kondo, Seiichi Kagoshima and Mitsuhiko Maesato,
Phys. Rev. B \textbf{67}, 134519 (2003).
\end{thebibliography}
\end{document}